\UseRawInputEncoding
\documentclass[lettersize,journal]{IEEEtran}
\usepackage{enumitem}
\usepackage{amsmath,amsfonts}
\usepackage{algorithmic}
\usepackage{algorithm}
\usepackage{array}
\usepackage[caption=false,font=normalsize,labelfont=sf,textfont=sf]{subfig}
\usepackage{textcomp}
\usepackage{stfloats}
\usepackage{url}
\usepackage{verbatim}
\usepackage{graphicx}
\usepackage{cite}
\usepackage{svg}
\usepackage{multirow}
\usepackage{array}

\usepackage{amssymb}
\usepackage{pifont}
\usepackage{xcolor}

\begin{document}

\title{Impact of Code Context and Prompting Strategies on Automated Unit Test Generation with Modern General-Purpose Large Language Models}

\author{Jakub Walczak, Piotr Tomalak and Artur Laskowski
\thanks{J. Walczak and P. Tomalak are affiliated with Lodz University of Technology, Łódź, Poland}
\thanks{A. Laskowski is senior software developer affiliated with Comarch, Łódź, Poland.}
}

\markboth{Journal of \LaTeX\ Class Files,~Vol.~14, No.~8, August~2021}%
{Shell \MakeLowercase{\textit{et al.}}: A Sample Article Using IEEEtran.cls for IEEE Journals}


\maketitle

\begin{abstract}
Generative AI is gaining increasing attention in software engineering, where testing remains an indispensable reliability mechanism. According to the widely adopted testing pyramid, unit tests constitute the majority of test cases and are often schematic, requiring minimal domain expertise. Automatically generating such tests under the supervision of software engineers can significantly enhance productivity during the development phase of the software lifecycle.

This paper investigates the impact of code context and prompting strategies on the quality and adequacy of unit tests generated by various large language models (LLMs) across several families. The results show that including docstrings notably improves code adequacy, while further extending context to the full implementation yields definitely smaller gains. Notably, the chain-of-thought prompting strategy --- applied even to 'reasoning` models --- achieves the best results, with up to 96.3\% branch coverage, a 57\% average mutation score, and near-perfect compilation success rate. Among the evaluated models, M5 (Gemini 2.5 Pro) demonstrated superior performance in both mutation score and branch coverage being still in top in terms of compilation success rate.

All the code and resulting test suites are publicly available at \url{https://github.com/peetery/LLM-analysis}.
\end{abstract}


\section{Introduction}

\IEEEPARstart{R}{ecent} years have brought significant advancements in artificial intelligence (AI), particularly in the areas of performance and productivity enhancement. However, AI --- and particularly large language models (LLMs) --- still suffer from several weaknesses. Among them, convincing but senseless content generation ('hallucination`), safety misalignment ('ethicality`) \cite{yang2024harnessing}, unfairness \cite{dai2024bias}, and limited processing context are the most critical.

In spite of these restrictions, and bearing in mind the limited and merely apparent creativity of LLMs \cite{chakrabarty2024art}, they have become versatile tools already widely used across a variety of domains (creative industries \cite{epstein2023art}, entertainment, reporting, and software engineering \cite{edison2021comparing} are just cases in point) for multiple tasks. Yet, in most pursuits, the output of LLMs should undergo strict human supervision, especially in activities related to privacy, security, health, or society. This is a gap filled by \textbf{hybrid intelligence} \cite{dellermann2019hybrid}, understood as amplifying human capabilities \cite{jarrahi2022artificial} and, in particular, productivity \cite{noy2023experimental} in schematic and tedious activities.

In this paper, we present a comprehensive survey in the context of hybrid intelligence in automated unit-test generation using general-purpose models (across a variety of LLM families) verified against different prompting strategies. We juxtapose them with the test cases of the software development practitioner.
We established and justified a solid methodology and, based on the results, proposed a set of guidelines for the process of automated efficient and reliable unit-test generation. We investigated several different approaches, elaborated and justified in detail in the Methodology section.

\section{Related works}
\subsection{Large Language Models}

Large language models, such as GPT, Claude, and Gemini, have recently gained significant attention in the field of software engineering due to their ability to understand and generate source code. These models are typically based on transformer architectures, which process input as a sequence of tokens using self-attention mechanisms \cite{vaswani2017attention}. Each output token is generated based on the entire input context, enabling the model to capture complex dependencies within the data.

A core concept in LLMs is the \textbf{token} mentioned before. It is the smallest chunk of text the model ingests and emits. Depending on the tokenizer (algorithm for converting text into tokens), a token might be an individual character, a frequent word, or a subword piece (e.g.~the word 'tokenization` split into 'token` + 'ization`). Subword-level tokenization such as Byte-Pair Encoding (BPE) \cite{sennrich2016neural} or WordPiece \cite{wu2016google} which strikes a balance between fixed vocabularies and open-ended text, allowing models to handle rare or compound words gracefully. During both training and inference, the transformer treats each token as an input embedding and applies self-attention across the entire sequence, generating outputs one token at a time conditioned on all preceding tokens. 
Besides pure processing meaning, tokens play significant role also in the usage cost of LLMs. The price of a single token varies significantly among LLM models (see Tab. \ref{tab:tokencost}) families and their 'advancement`.

\begin{table}[ht]
\caption{Differences in prices of input tokens for chosen LLMs}
\label{tab:tokencost}
\centering
\begin{tabular}{|c||c|}
\hline
\textbf{Model} & \textbf{Cost per $10^6$ tokens [\$]}    \\ \hline
GPT-4.1 mini & $0.40$ \\ \hline
GPT-4.1 & $2.00$ \\ \hline
Claude Haiku 3 & $0.25$ \\ \hline
Claude Opus 4 & $15.00$ \\ \hline
\end{tabular}
\end{table}

In conversational settings, such as with ChatGPT, the full conversation history is included as a part of the prompt context at each step. This means that LLMs operates \textbf{statelessly} --- they do not 'remember` previous interactions unless those interactions are explicitly included in the current input. \textbf{As such, the complete context visible to the model is the concatenation of all prior messages in the conversation window}, subject to the model's token limit \cite{yang2024harnessing}.

This design makes \textbf{prompt engineering} --- the deliberate crafting and structuring of prompts --- critical for eliciting accurate, relevant outputs. Even minor violations in phrasing or the order of context examples can significantly affect LLM performance, especially in structured tasks such as code completion or test-case generation \cite {wei2022chain, chakrabarty2024art}.

Despite their general capabilities, LLMs can suffer from limitations such as hallucinations (confidently incorrect outputs) \cite{huang2025survey}, invalid syntax, or shallow semantic understanding. These issues are especially problematic in automated software testing, where both syntactic correctness and logical coverage are of utmost importance \cite{dai2024bias}.

\subsection{Testing in software development process}
Software testing is one of the most critical phases of the Software Development Life Cycle as it increases reliability of software (and hence, a company) and enable to reduce potential costs wreaked by buggy releases \cite{gupta2022study}. Testing can be categorized into three main techniques: 
\begin{itemize}
    \item functional,
    \item performance,
    \item security.
\end{itemize}

During the coding phase, developers primarily focus on verifying the functionality of their code, which falls under the category of functional testing. This category includes, among others, \textbf{unit tests} and \textbf{integration tests}.

The type and number of tests required in a project depend on the nature of the software. Projects with complex domain logic --- often among the most profitable --- tend to require a large number of tests to verify the correctness of that logic. Domain logic is independent of technologies or frameworks; rather, it represents the business rules behind the process. For instance, domain logic might define whether an order can transition directly from the \texttt{PENDING} state to the \texttt{APPROVED} state, or if it must first pass through the \texttt{PAID} state. Unit tests are typically used to verify such scenarios.

For systems with rich domain logic, the well-known testing pyramid approach introduced by Cohn \cite{cohn2010succeeding} is commonly adopted. This framework, further discussed in \cite{contan2018test,mukhin2021testing}, emphasizes the importance of unit tests, which usually form the majority of a project's test suite and occupy a significant portion of developers' time. This is particularly true for tests that do not require deep domain expertise but still need to be written exhaustively.

These factors motivate our focus on automated unit test generation. However, unit tests should not be evaluated solely based on quantity; their quality --- such as the uniqueness and relevance of the scenarios they cover --- is equally important.

\subsection{Automation of Software Engineering Processes}
Automation plays a crucial role across the entire Software Development Life Cycle (SDLC). It has been pivotal in boosting productivity, reducing costs, and improving the quality of IT systems by enabling repeatable, transparent workflows with minimal susceptibility to human error \cite{akbar2022toward}, at least for classical rule-based automation.

From early phases such as modelling and requirements specification \cite{orlova2013approach}, through development and testing \cite{huttermann2012devops,schafer2023empirical,yuan2023evaluating}, to documentation \cite{bhatia2018ontology}, nearly every stage of the SDLC is amenable to automation. This trend has become particularly prominent in the era of generative artificial intelligence, which enable new generative-based automation paradigm. However, it is important to note that automation in software engineering predates the rise of large language models (LLMs). The broader DevOps ecosystem has long pursued the goal of automating software delivery pipelines to enhance productivity in development, deployment, and monitoring \cite{luz2019adopting,akbar2022toward}.

Automation is especially critical for delivering high-quality software by minimizing human involvement in repetitive and error-prone tasks. Yet, when using generative AI for tasks such as coding and testing, human oversight remains essential. While generative models are deterministic systems, they lack full transparency and predictability. As such, automation involving generative AI --- including LLMs --- requires additional attention and rigour to ensure reliability and correctness.

\section{Methodology}
Developing a reliable methodology for evaluating the effectiveness of automated unit test generation is challenging, primarily due to the lack of established benchmarks and the absence of objective criteria for assessing test adequacy --- that is, the extent to which generated tests reflect real-world development needs. This issue is evident in benchmarking efforts such as those presented in \cite{jain2024livecodebench} and \cite{wang2025testeval}.

Furthermore, publicly available datasets pose an additional difficulty: the risk of data leakage. As noted in \cite{walczak2025evidence}, details of datasets used during the training of large language models may overlap with those used for evaluation, wreaking performance assessments potentially unreliable.

To address these challenges, we propose an updated methodology for evaluating the effectiveness of LLMs in the task of automated unit test generation.

\subsection{Investigation subject}
Unlike previous studies, we simplified the subject of comparison to minimize the influence of confounding factors related to software development on the test case generation process. To this end, \textbf{we designed and implemented twelve custom Python methods} that focuses on core behaviours relevant to unit testing. 

We chose to conduct our experiments on this custom, previously unpublished code to mitigate the risk of so-called \textit{data leakage} \cite{kapoor2024reforms,walczak2025evidence}. By using code that has not been publicly released, we ensure that it was not part of the training data for any large language model (LLM). That makes result more creditable. We are aware, the size of the dataset is limited but it demonstrates well a single unit of code and future research will be focused on preparing and examining more custom pieces of code.

The test subject used in all experiments was a set of custom-developed Python methods, simulating a minimalistic shopping cart system. Although the application does not demonstrate the complexity of real-world IT systems, it serves as a valid subject of investigation. The results can be generalized to more complex systems, as the selected subject can be seen as an individual component of a larger software architecture, which fully aligns with the paradigm of unit testing~\cite{ellims2004unit}.

The code was intentionally crafted to represent a realistic but isolated code component. It exposes a clearly defined public API and implements various functionalities commonly found in production systems: item management (addition and removal), subtotal computation, tax and discount application, shipping logic, and calculation of the final total. The diversity of logic paths makes it suitable for evaluating test coverage and the robustness of generated test cases.

To ensure compatibility with LLM-based test generation workflows, special attention was given to:
\begin{itemize}
    \item comprehensive documentation using detailed docstrings for each method,
    \item static typing through Python type hints to provide signature-level context,
    \item strict runtime validation: including type checks and well-defined exception raising for invalid or out-of-range inputs.
\end{itemize}

The design of the class facilitates controlled experimentation and repeatability. It enables analysis of how various levels of context (e.g., method signatures, docstrings, full implementation) influence the quality of test cases generated by large language models. In particular, the class supports diverse scenarios, including standard logic, input validation, and edge cases such as extreme numerical values or operations on large item lists.


\subsection{Quality measures}
\label{sec:qualitymeasures}

Following prior research in the area of automated unit-test generation \cite{yuan2023no,yang2024empirical,lops2025system}, we applied well-established quality measures to evaluate models and code features from different perspectives. 

It is yet important to note that the adequacy of unit tests encompasses more than the quantitative metrics described below. Many essential factors cannot be directly measured. These include alignment with project-specific conventions, adherence to used libraries, and consistency in assertion styles. Such qualitative aspects should be carefully evaluated during the deployment phase of LLM-based test generation systems.

\subsubsection{Syntactic correctness}
To evaluate the test suites from the code syntax perspective, we computed the run success rate (compilation success rate, CSR), expressed as the fraction of test cases executed without exception relative to all produced cases (\ref{eq:csr}).

\begin{equation}
    CSR = \frac{|S|}{|T|} \cdot 100\%,
    \label{eq:csr}
\end{equation}
where $S$ is the set of tests that ran without any exceptions, $T$ is the set of all unit tests, and $|\cdot|$ denotes the cardinality.

\subsubsection{Code coverage}
To determine the extent to which the code is exercised by the unit tests, we applied \textbf{branch coverage} (BC) --- which evaluates decision points in the code \cite{baluda2015bidirectional} ---and method coverage (MC), which verifies whether each method has at least one corresponding test case.

\subsubsection{Unit test detectability}
To assess the ability of the generated tests to detect logical changes in the code, we computed the \textbf{mutation score} (MS) \cite{fraser2010mutation}, expressed as the ratio of detected logical mutations (\textit{mutants}). We designed the mutants manually. The strategy and justification for each code mutation are presented in detail in Appendix \ref{app:mutants}.

\subsubsection{Test uniqueness}
To evaluate the actual number of unique scenarios covered by the test suite, we manually excluded duplicated test cases and compared the count of unique test scenarios (UT).

\subsubsection{Response generation time}
While the efficiency of response generation is not as critical as the adequacy of tests in the context of unit testing for IT systems, it remains an important factor for ensuring that development processes (including unit testing) are completed within manageable time bounds. This is essential for maintaining cost-effectiveness in IT companies. To address this aspect, we measured the time required for each model to generate the output test suite.

\subsection{Used LLM models}
In our experiments, we evaluated 6 large language models across well-established and referenced LLM families (Tab. \ref{tab:llms}):

\begin{table}[ht]
\caption{Large language models used in the experiments}
\label{tab:llms}
\centering
\begin{tabular}{|c||c|c|}
\hline
\textbf{Model} & \textbf{Family} & \textbf{Model}       \\ \hline
         M1    & OpenAI          &  GPT-4.5          \\ \hline
         M2    & OpenAI          &  GPT-o3          \\ \hline
         M3    & OpenAI          &  GPT-o4-mini-high   \\ \hline
         M4    & Anthropic       &  Claude 3.7 Sonnet   \\ \hline
         M5    & Google DeepMind &  Gemini 2.5 Pro     \\ \hline
         M6    & DeepSeek        &  DeepSeek-V3       \\ \hline
\end{tabular}
\end{table}

\subsection{Description style}
In our study, we evaluated two types of prompting strategies expressed in natural language:

\textbf{S1.} Simple prompting

\textbf{S2.} Chain-of-thoughts

Both strategies include the task scope, specifically: impersonation details, the task to be completed, technologies used, and the code feature (CF) being the code context considered by an LLM.

\textbf{Simple prompting} (S1) is a two-step interaction with the model:

\begin{tabular}{lp{7cm}}
\textbf{Step 1.} & Present the task scope \\ 
\textbf{Step 2.} & Request unit-test implementation for the given code feature. \\
\end{tabular}

The other strategy, widely referred to as \textit{chain-of-thought} (S2), is a common prompting technique \cite{wei2022chain} based on breaking down a complex task into intermediate (simpler) steps that collectively produce the desired output. In our case, it is implemented as a three-step interaction with the LLM:

\begin{tabular}{lp{7cm}}
\textbf{Step 1.} & Present the task scope \\ 
\textbf{Step 2.} & Request generation of test scenarios (test cases) \\
\textbf{Step 3.} & Request unit-test implementation for the given code feature. \\
\end{tabular}

\subsection{Code features}
We adapted a set of commonly used code features to evaluate different levels of code context details considered by an LLM:

\begin{tabular}{lp{7cm}}
\textbf{CF1.} & Method signatures (with type hints) \cite{yang2024empirical}, \\
\textbf{CF2.} & Method signatures and docstrings, \\
\textbf{CF3.} & Complete method implementation with docstrings. \\
\end{tabular}

\subsection{Experiment procedure}
To ensure reliable and reproducible evaluation, all experiments were conducted in a controlled setting, with strict prompt consistency and isolation of context. The goal was to assess how different levels of contextual information influence the effectiveness of unit test generation across several large language models.

For each combination of the prompting strategy (S1, S2) and the code context level (CF1-CF2) defined in the previous section, we constructed a single prompt following the same pattern. The prompt specified the expected task, constraints, framework (unittest), output format, and included only the relevant code context (method signatures, method signatures and docstrings, or complete method implementation with docstrings) depending on the current variant.

To ensure a clean inference environment and prevent contextual leakage between runs, a new chat session was initialized for each model-prompt pair. Thus, the only context available to the model was the provided prompt. The prompt content remained exactly the same across different models for a given condition, which allowed for consistent cross-model comparison.

Repeatability of LLM models is usually tuned with the \textit{temperature} parameter. We have used the default value to ensure lack of variability in the output.

In the case of models offering configuration options (e.g., temperature), default or zero-temperature settings were used when accessible to reduce randomness and emphasize deterministic behaviour, aligning with prior recommendations in prompt-based testing evaluations~\cite{yuan2023no}.

Each prompt was carefully crafted in line with its respective strategy:
\begin{itemize}
    \item For \textbf{Simple prompting (S1)}: a single prompt included all instructions and code context.
    \item For \textbf{Chain-of-thought prompting (S2)}: a sequence of three messages guided the model through reasoning steps, scenario analysis, and final test generation.
\end{itemize}

All LLM outputs were then passed through the evaluation pipeline described in Section~\ref{sec:qualitymeasures}, including scenario comparison, syntactic validation, coverage measurement, and detectability. The metrics were collected in tabular format and compared to identify strengths and weaknesses in the models and strategies.

\section{Results \& Discussion}
The exhausting results of experiments we carried out are presented in the Table \ref{tab:results}). The summary grouped by prompting strategy and code features used are, in turn, presented in Table \ref{tab:summary1}.

\begin{table*}[]
\centering
\caption{The results of experiments for two versions of prompting strategy and three types of code features. Results for all evaluated LLM models against the results for tests generated by code practitioner (SP)}
\label{tab:results}
\begin{tabular}{|c|c|c||c|c|c|c|c|}
\hline
\textbf{Promp.} & \textbf{CF} & \textbf{Model} & \textbf{\# tests} & \textbf{CSR [\%]} & \textbf{MS [\%]} & \textbf{BC [\%]} & \textbf {Time [s]} \\ \hline
\multirow{18}{*}{S1} & \multirow{6}{*}{CF1} & M1 & 17 & 88 & 43 & 63 & 61\\
\cline{3-8} 
& & M2 & 22 & 86 & 48 & 68 & 41\\
\cline{3-8}
& & M3 & 33 & 88 & 48 & 84 & 20\\
\cline{3-8}
& & M4 & 40 & 74 & 36 & 68 & 46\\
\cline{3-8}
& & M5 & 80 & 85 & 51 & 96 & 40\\
\cline{3-8} 
& & M6 & 23 & 78 & 41 & 64 & 54\\
\cline{2-8}
& \multirow{6}{*}{CF2} & M1 & 26 & 96 & 41 & 80 & 75\\
\cline{3-8}
& & M2 & 45 & 100 & 50 & 89 & 35\\
\cline{3-8}
& & M3 & 30 & 97 & 50 & 98 & 30\\
\cline{3-8}
& & M4 & 47 & 100 & 50 & 98 & 68\\
\cline{3-8}
& & M5 & 73 & 64 & 59 & 98 & 60\\
\cline{3-8}
& & M6 & 34 & 97 & 47 & 98 & 71\\
\cline{2-8}
& \multirow{6}{*}{CF3} & M1 & 24 & 100 & 47 & 68 & 97\\
\cline{3-8}
& & M2 & 36 & 100 & 49 & 89 & 72\\
\cline{3-8}
& & M3 & 51 & 100 & 50 & 98 & 39\\
\cline{3-8}
& & M4 & 46 & 100 & 50 & 98 & 46\\
\cline{3-8}
& & M5 & 72 & 100 & 86 & 98 & 71\\
\cline{3-8}
& & M6 & 42 & 95 & 41 & 98 & 82\\
\cline{1-8}
\multirow{18}{*}{S2} & \multirow{6}{*}{CF1} & M1 & 40 & 88 & 50 & 71 & 176\\
\cline{3-8}
& & M2 & 56 & 77 & 50 & 77 & 111\\
\cline{3-8}
& & M3 & 64 & 69 & 48 & 84  & 113\\
\cline{3-8}
& & M4 & 60 & 68 & 43 & 82 & 147\\
\cline{3-8}
& & M5 & 95 & 83 & 53 & 98 & 155\\
\cline{3-8}
& & M6 & 37 & 76 & 44 & 70 & 125\\
\cline{2-8}
& \multirow{6}{*}{CF2} & M1 & 43 & 98 & 47 & 86 & 245\\
\cline{3-8}
& & M2 & 58 & 98 & 51 & 98 & 112\\
\cline{3-8}
& & M3 & 65 & 100 & 52 & 98 & 85\\
\cline{3-8}
& & M4 & 75 & 99 & 52 & 98 & 143\\
\cline{3-8}
& & M5 & 88 & 69 & 69 & 98 & 182\\
\cline{3-8}
& & M6 & 53 & 96 & 46 & 96 & 177\\
\cline{2-8}
& \multirow{6}{*}{CF3} & M1 & 40 & 98 & 47 & 88 & 270\\
\cline{3-8}
& & M2 & 49 & 100 & 53 & 98 & 138\\
\cline{3-8}
& & M3 & 67 & 100 & 54 & 98 & 116\\
\cline{3-8}
& & M4 & 56 & 100 & 50 & 98 & 175\\
\cline{3-8}
& & M5 & 96 & 100 & 87 & 98 & 170\\
\cline{3-8}
& & M6 & 61 & 100 & 50 & 98 & 200\\
\hline
 \multicolumn{3}{|c||}{SP} & 54 & 100 & 44 & 98 & - \\ \hline
\end{tabular}
\end{table*}

\begin{table*}[h]
\centering
\caption{Summary statistics grouped by prompting strategy and code features used}
\label{tab:summary1}
\begin{tabular}{|c|c|c|c|c|c|c|}
\hline
\textbf{S} & \textbf{CF} & \textbf{\# tests} & \textbf{CSR [\%]} & \textbf{MS [\%]} & \textbf{BC [\%]} & \textbf{Time [s]} \\
\hline
\multirow{3}{*}{S1} 
  & CF1 & $36 \pm 21$ & $83.17 \pm 5.30$ & $44.50 \pm 5.06$ & $73.83 \pm 12.09$ & $44.33 \pm 15.04$\\ \cline{2-7}
  & CF2 & $43 \pm 16$ & $92.33 \pm 12.76$ & $49.50 \pm 5.32$ & $93.50 \pm 6.87$ & $54.83 \pm 18.84$\\ \cline{2-7}
  & CF3 & $45 \pm 15$ & $99.17 \pm 1.86$ & $53.83 \pm 14.71$ & $91.50 \pm 11.01$ & $67.83 \pm 22.95$\\
\hline
\multirow{3}{*}{S2} 
  & CF1 & $59 \pm 19$ & $76.83 \pm 7.10$ & $48.00 \pm 3.51$ & $80.33 \pm 9.43$ & $137.17 \pm 25.76$\\ \cline{2-7}
  & CF2 & $64 \pm 15$ & $93.33 \pm 10.95$ & $52.83 \pm 7.60$ & $95.67 \pm 4.38$ & $159.83 \pm 60.44$\\ \cline{2-7}
  & CF3 & $62 \pm 18$ & $99.67 \pm 0.75$ & $56.83 \pm 13.68$ & $96.33 \pm 3.73$ & $173.17 \pm 55.02$\\
\hline
\end{tabular}
\end{table*}

Based on the experiments results analysis, several key observations emerged regarding the impact of prompting strategies and code context levels on unit test generation quality across different LLM models.

\subsection{Impact of Code Context Level}

The analysis reveals a clear pattern demonstrating the influence of available code context on test generation quality. This aligns with the intuition. Full context (CF3) --- including the docstring and full implementation --- consistently yields the highest performance across most evaluation metrics. Models with access to the complete implementation achieve higher CSR, improving by an average of 6.84 percentage points (pp) for S1 and 6.34pp for S2, compared to models with access only to the function signature and docstring (CF2). Mutation scores also show a notable improvement, exceeding CF2 by 4.33pp for S1 and 4.00pp for S2. Under the chain-of-thought (S2) prompting strategy, branch coverage reaches its highest value for CF2 (96.33\%), whereas with the simple prompting strategy (S1), the branch coverage is approximately 2pp lower compared to CF2. The advantages of CF3 over the minimal context configuration (CF1) are even more pronounced across all evaluated metrics.

In general, the results demonstrate increase in quality measures when using chain-of-thoughts prompting comparing to the simple prompting, however, that increase is not as significant as we could expect and aggregated differences are statistically negligible.

It is important to note that the tested methods contain a single code branch that could not be covered through the class's public interface. This branch represented an exception condition in the main calculation method that was prevented by validation logic in another method. Consequently, the theoretical maximum branch coverage was 98\%. Notably, no LLM attempted to exceed the prescribed interface boundaries to force coverage of this unreachable code segment, demonstrating appropriate adherence to proper testing practices by focusing on the class's intended interface rather than employing artificial manipulation techniques. It is important observations.

\subsection{Differences Between Prompting Strategies}
Chain-of-thought prompting (S2) generally surpasses simple prompting (S1) in the number of generated tests and scenario coverage, though this does not consistently translate to superior quality. Models employing S2 generate approximately 20-40\% more tests, but this frequently results in higher proportion of 'redundant` tests that duplicate scenario coverage.

A notable exception is model M1 (GPT-4.5), which, with chain-of-thought and full context, generates only 40 tests at 98\% CSR, while simple prompting produces 24 tests at 100\% CSR, suggesting that increased reasoning  can occasionally lead to model confusion. 

\subsection{Individual Model Characteristics}

M1 (GPT-4.5) demonstrates the most inconsistent performance among all evaluated models, with coverage ranging dramatically from 76-91/\% and consistently the lowest efficiency in test generation. This model exhibits significant context confusion, particularly struggling with chain-of-thought prompting where it often describes more test scenarios than it actually implements. Despite generating 24-43 tests, it typically covers only 22-37 scenarios, indicating poor scenario-to-test efficiency. Its primary strength lies in producing clean, well-organized test structures with atomic design principles, though it consistently requires the longest generation time and frequently misses critical corner cases ad type validation tests.

M2 (GPT-o3) establishes itself as a highly consistent performer, maintaining excellent 99\% statement coverage across most experimental conditions while achieving solid 48-53\% mutation scores. This model demonstrates remarkable efficiency, covering 46-48 scenarios with relatively few tests (19-49), indicating superior test design capabilities. However, it exhibits notable context sensitivity, with performance degrading significantly in interface-only scenarios. GPT-o3 excels in edge case detection and maintains clean test organization, though it consistently lacks robustness testing for extreme values and performance evaluation capabilities.

M3 (GPT-o4-mini-high) stands out as the most reliable model with 100\% compilation success rate and consistently excellent 99\% statement coverage. It achieves the highest mutation scores (50-54\%) among GPT variants while generating comprehensive test suites (51-67 tests). This model uniquely in corporates extreme value testing, including calculations with numbers as large as $10^6$, demonstrating superior edge case exploration. Its comprehensive approach covers 48 scenarios typically, though this thoroughness occasionally results in test redundancy. 

M4 (Claude 3.7 Sonnet) exhibits highly context-dependent behaviour, ranging from exceptional performance (99\% coverage) with full context to poor results (87\% coverage) with limited context. This model frequently demonstrates instruction non-compliance, often ignoring explicit prompts about avoiding code comments or implementation details. Uniquely among all models, Claude consistently tests None value handling and occasionally attempts to implement its own \texttt{OrderCalculator} class rather than testing the provided one. When operating with full context and simple prompting, it achieves perfect 1:1 test-to-scenarios ratios, indicating optimal efficiency under ideal conditions.

M5 (Gemini 2.5 Pro) distinguishes itself as the superior performer regarding mutation score, achieving exceptional values (86-87\% with simple prompting and full context). This model also demonstrates remarkable creativity in generating precision calculation tests and implements advanced techniques such as floating-point comparison methods. However, it exhibits significant reliability variability, with compilation success rates ranging from 64-100\% depending on context availability. Gemini generates the most comprehensive test suites (47-96 tests), occasionally producing excessive similar tests that may indicate over-engineering tendencies. Notably, this model show pronounced context sensitivity, experiencing hallucination problems when working with limited context scenarios, yet achieves unmatched performance when provided with complete implementation details.

M6 (DeepSeek) emerges as the most efficient model, consistently achieving 97-99\% statement coverage with exceptional test-to-scenario ratios. It demonstrates remarkable consistency across all experimental conditions while maintaining the fastest generation times. DeepSeek generates precisely what is needed (34-61 tests) with minimal redundancy, often covering 46 scenarios with optimal efficiency. However, its conservative approach may limit innovative test scenario discovery, and it consistently lacks comprehensive edge case exploration compared to more creative models like Gemini.

\subsection{Scenario Coverage Observations}

The analysis reveals systematic gaps across all models. Performance analysis (e.g., adding thousands of products), frequently being part unit testing \cite{horky2013performance,horky2015utilizing}  are consistently omitted by all models regardless of configuration. Similarly, robustness tests with values such as \texttt{None}, \texttt{inf}, \texttt{NaN} are rarely implemented.

Precision calculation tests are implemented selectively...

\subsection{Response Generation Time Impact}

The experimental results reveal significant variations in response generation time across different prompting strategies and model configurations. Simple prompting (S1) demonstrates substantially faster generation times, ranging from 20 to 97 seconds, with a mean response time approximately 55 seconds. In contrast, chain-of-thought prompting (S2) exhibits considerably longer generation times, spanning from 85 to 270 seconds, with a mean response time of approximately 158 seconds. It represents 187\% increase compared to simple prompting. That follows the intuition as chain-of-thought require processing more tokens than simple prompting.

The analysis indicates that code feature complexity directly correlates with generation time. Interface-only context (CF1) generally produces the fastest responses, while full context (CF3) typically requires the longest processing time. This pattern suggests that increased contextual information, while potentially improving test quality metrics, introduces computational overhead that significantly impacts response latency. That should be considered when adapting LLM in software development life cycle.

Model-specific performance varies considerably with M3 consistently demonstrating the fastest response times across most configurations, while M1 and M6 tend to exhibit longer generation times, particularly in complex prompting scenarios. The temporal performance differences between models highlight the importance of model selection in production environments where response time constraints are critical.

\subsection{Compare with Software Practitioner Test Suite}

The software partitioner (SP) baseline provides a crucial reference point for evaluating LLM-generated test suite quality. The human-authored test suite comprises 54 test cases with perfect compilation success rate (100\%), moderate mutation score (44\%), and excellent branch coverage (98\%). This baseline establishes the gold standard for syntactic correctness while revealing areas where automated approaches may excel. 

Comparative analysis reveals that LLMs frequently surpass the practitioner baseline in test quantity, with several configurations generating 60-96 tests cases compared to the practitioner's 54. However, this quantitative advantage often comes with trade-offs in compilation reliability, as evidenced by variable compilations success rates across LLM configurations.

Notably, many LLM configurations achieve superior mutation scores compared to the practitioner baseline, with some reaching 50-87\% versus the practitioner's 44\%. This suggests that automated test generation can identify edge cases and fault conditions that may be overlooked in manual test development.

The comparison reveals a fundamental trade-off between syntactic reliability and comprehensive fault detection. While the practitioner ensures perfect compilation through domain expertise and iterative refinement, LLMs demonstrate superior capability in generating diverse test scenarios that expose subtle defects through mutation testing.

\section{Limitations}
We should think of LLMs as ordinary yet versatile tools and every tool has its limits within which it is productive and useful. During our research, we identified a few of them in the process of automated unit tests generation, some of which align with the common limitations of large language models.

\begin{description}
    \item[Syntactic Reliability Inconsistency] -- Our experimental results demonstrate significant variability in compilation success rates across different model configurations, ranging from 64\% to 100\%. This inconsistency poses challenges for production deployment, as developers cannot reliably predict whether generated test suites will compile without manual intervention. The variation appears to be model-dependent rather than systematic, making it difficult to establish consistent quality assurance protocols.
    \item[Context Processing Limitations] -- The analysis reveals that models struggle with optimal context utilization. While increased code context (CF3) generally improves test coverage metrics, it also substantially increases response generation time and occasionally degrades compilation success rates. This suggests that current LLMs have difficulty efficiently processing and leveraging comprehensive code context without introducing syntactic errors or computational overhead.
    \item[Scenario Coverage Gaps] -- Despite generating substantial numbers of test cases, all evaluated models consistently fail to address critical testing scenarios. Performance testing scenarios and robustness testing with edge values such as \texttt{None}, \texttt{inf}, and \texttt{NaN} are systematically omitted across all configurations. This limitation indicates a fundamental gap in understanding comprehensive testing requirements beyond basic functional coverage.
    \item[Prompt Engineering Sensitivity] -- The dramatic difference in response generation time between simple prompting (20-97s) and chain-of-thought prompting (85-270s) demonstrates excessive sensitivity to prompt formulation. This 187\% average increase in processing time for marginally improved test quality suggests that current models lack efficient reasoning pathways and are highly dependent on prompt engineering expertise.
    \item[Model Selection Unpredictability] -- The substantial performance variations between models (M1-M6) across identical tasks reveal the challenge of optimal model selection. Without extensive empirical evaluation, practitioners cannot predict which model will perform optimally for specific testing scenarios, limiting the practical applicability of automated test generation in diverse software development contexts.
\end{description}

\section{Ethical Considerations}

\subsection{Human Oversight}
Despite the impressive advancements in artificial intelligence, it is crucial to remember that AI systems --- including large language models --- are not conscious entities and cannot autonomously validate their outputs. All AI-generated content requires human oversight, with no exceptions. While certain applications, such as digital art generation, may tolerate a looser approach to supervision, software development --- especially in domains involving personal or sensitive data or decision-critical systems --- demands rigorous and responsible oversight.

\subsection{Responsibility}
Another key ethical issue is responsibility. AI, no matter how advanced, lacks moral awareness. It is ultimately a computational tool, a sequence of mathematical operations without intention or accountability. Therefore, responsibility for any AI-generated output rests solely with the human operator supervising its use. In the context of unit test generation, the validity and adequacy of the produced tests are the sole responsibility of the software developer or organisation which should develop solid and systematic code-review pipelines to ensure the right quality of produced software.

\subsection{Biased Output}
All LLMs are inherently biased due to the datasets used during their training phase. These biases are never negligible and should be considered carefully when using LLMs for automated unit test generation. LLMs may unintentionally overlook important edge cases or incorrectly classify them as unnecessary. Moreover, large language models can follow stereotypes and reason schematically (e.g. variable naming), which is not always desirable in the process of software development. Therefore, LLMs should be carefully tuned to align with the specifications of a particular software project and, let us emphasize it again, their output must always be reviewed and guided by a human expert.

\subsection{Intellectual Property}
The capabilities of LLMs derive largely from the data used during their training. This tight coupling between data and model performance is characteristic of all machine learning systems. Consequently, LLM providers often refrain from disclosing comprehensive details about the datasets used in training. This lack of transparency raises important concerns regarding intellectual property, particularly the unauthorized use or reproduction of proprietary code found in public or scraped data.

These concerns are especially relevant in software development, where even partial reproduction of licensed code may create a copyright violation. Some LLM providers, such as Microsoft (provider of the GitHub Copilot tool), have acknowledged this risk. They have implemented copyright policies and publicly committed to assuming responsibility for copyright violations caused by their tools \cite{smith2023microsoft}.

Intellectual property remains a crucial ethical aspect to consider when integrating LLM-based tools into software development pipelines, especially in commercial or proprietary projects.


\section{Conclusions}
In our paper, we analyse several general-purpose large language model in the task of automated unit test generation in varying width of code context in two prompting strategies. 

Based on the result, we have drawn several conclusions. Primarily, general-purpose models, especially M5, is superior in the mutation score across all code context and prompting strategies producing, at the same time, the much more (than other models) test cases covering the most of code branches. Those makes it the best choice for automated unit-tests generation.

The code context do matter. The widest investigated context including method's signature and body (implementation) yielded the best results. For CF3 mutation score was the highest in both S1 and S2 prompting strategy, meaning provided code implementation does produce only 'happy tests` but cases that truly verify the correctness of logic. That is critical.

The results we collected challenge the commonly held assertion by LLM authors that chain-of-thought prompting can degrade the performance of reasoning models (specifically, models with self-consistency mechanism \cite{wang2022self}). On the contrary, our findings clearly demonstrate that applying this strategy (even for 'reasoning models`) leads to improved outcomes in test generation tasks. On the other hand, that increase in quality measures is, on average, not impressive and the users of LLM should deeply analyse the cost-quality trade-off as extended prompting strategy is bonded to higher usage costs due need to process more tokens.

\section*{Authors' contribution}
\begin{enumerate}
    \item Conceptualization:  J.W., P.T.;
    \item Methodology: J.W.;
    \item Implementation: P.T.;
    \item Formal Analysis: J.W., P.T.;
    \item Investigation: P.T.;
    \item Writing - Original Draft Preparation: P.T.,J.W., A.L.;
    \item Writing - Review \& Editing:  J.W., P.T., A.L.;
    \item Supervision: J.W.
\end{enumerate}

\section*{Impact Statement}
We expect that the research and analysis presented in this paper will contribute to a more sceptical and informed use of large language models for automated unit test generation, particularly in the context of improving software developer productivity. We also highlight several ethical concerns that must be considered when integrating LLMs into software development workflows.

Furthermore, we anticipate that the performance metrics reported --- obtained using a rigorous and transparent methodology --- will help practitioners and researchers make more informed choices when selecting from among the evaluated LLM families and models.

We hope this work will also motivate future research into safer, more interpretable LLM-based tools for software testing.

\section*{Generative AI statement}
In out studies, we used generative AI in two ways. First of all, different generative models were used as investigation subjects in our research ---- details on selected models are presented in the Methodology section. We also used generative AI, in particular ChatGPT (GPT-4o) model only to fix language error and polish style.

\section*{Supplementary Materials}
All code analysed in this study, along with the prompts used and the corresponding outputs, are provided as supplementary materials. These resources are publicly available at: \url{https://github.com/peetery/LLM-analysis}.

{\appendices
\section{Generation of mutants for mutation analysis}
\label{app:mutants}

To evaluate the robustness of unit tests generated by large language models, we employed mutation testing using the \texttt{mutmut} framework \cite{mutmut_docs}. Mutation testing involves introducing small changes, known as \textit{mutants}, into the source code to simulate potential faults. The effectiveness of a test suite is then assessed based on its ability to detect these mutations (\textit{detectability}).

\texttt{mutmut} systematically applies a variety of mutations to the codebase, such as altering arithmetic operators, modifying logical conditions, and changing constant values. After each mutation, the existing test suite is executed. If a test fails, the mutant is considered 'killed`, indicating that the test suite successfully detected the introduced fault. Conversely, if all tests pass, the mutant 'survives`, suggesting a potential gap in the coverage of tested scenarios.

For our experiment, we configured \texttt{mutmut} to target the investigated methods. The test suites generated by different LLMs were then evaluated against these mutants. The proportion of killed mutants to the total number of mutants provides the \textit{mutation score}, a quantitative measure of the test suite's fault detection capability.

This approach allowed us to systematically asses and compare the effectiveness of LLM-generated test suites in identifying potential faults within the codebase.
}

\bibliographystyle{IEEEtran}
\bibliography{IEEEabrv,references}


\section*{Biography Section}

\begin{IEEEbiography}[{\includegraphics[width=1in,height=1.25in,clip,keepaspectratio]{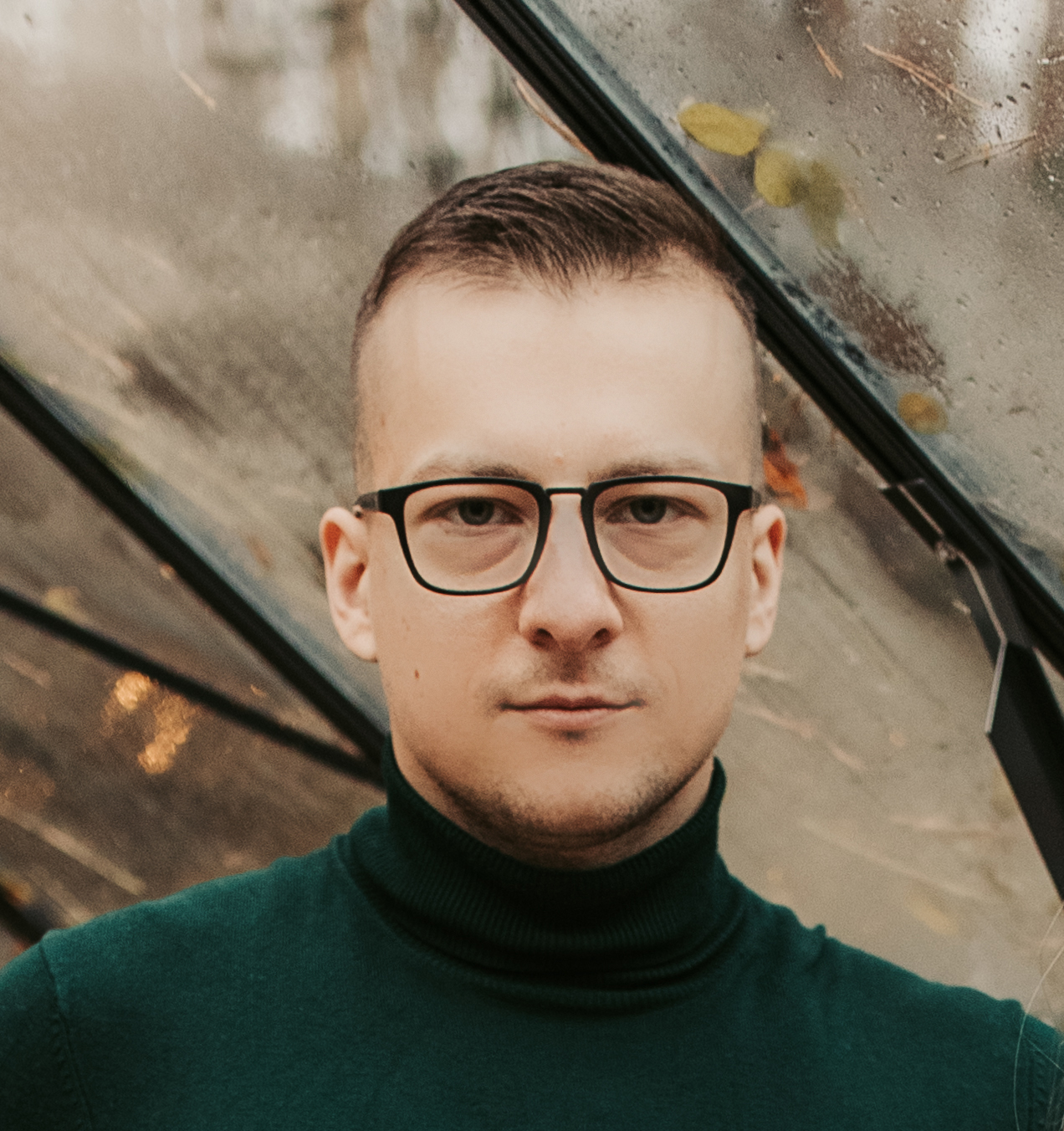}}]{Jakub Walczak}
defended his PhD in 2022, he is
an AI researcher with both academic and industrial
experience across a wide range of domains, including climate and ICT industry. He specializes in the
ethical and multidisciplinary applications of artificial intelligence, bio-inspired advancements, and the development of innovative approaches for AI.
\end{IEEEbiography}

\begin{IEEEbiography}[{\includegraphics[width=1in,height=1.25in,clip,keepaspectratio]{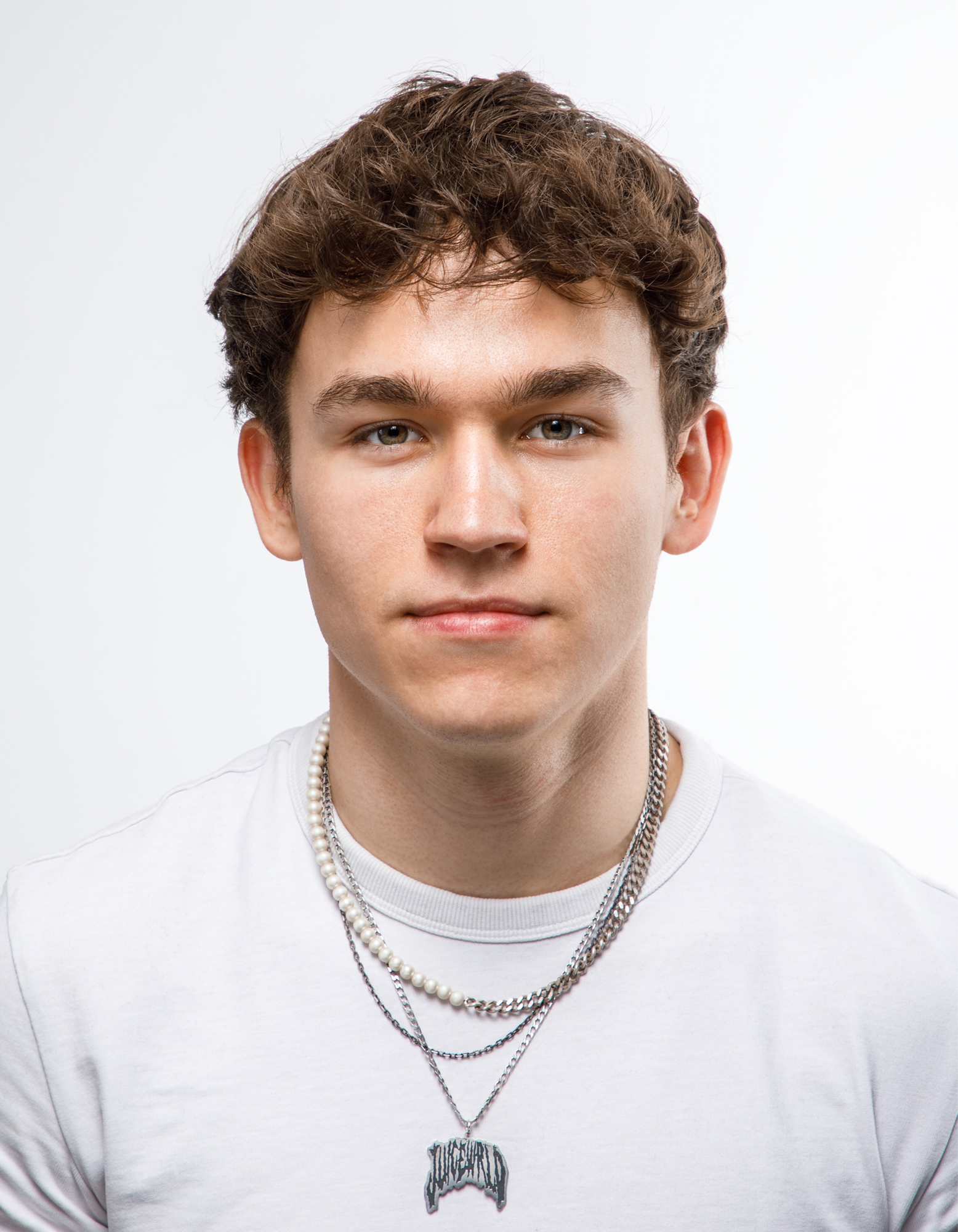}}]{Piotr Tomalak}
is a Bachelor's student at the Faculty of Technical Physics, Information Technology and Applied Mathematics at Łódź University of Technology. As part of his engineering thesis, he is conducting research focused on the use of artificial intelligence --- specifically large language models in the automatic generation and evaluation of unit tests in software engineering.
\end{IEEEbiography}

\begin{IEEEbiography}[{\includegraphics[width=1in,height=1.25in,clip,keepaspectratio]{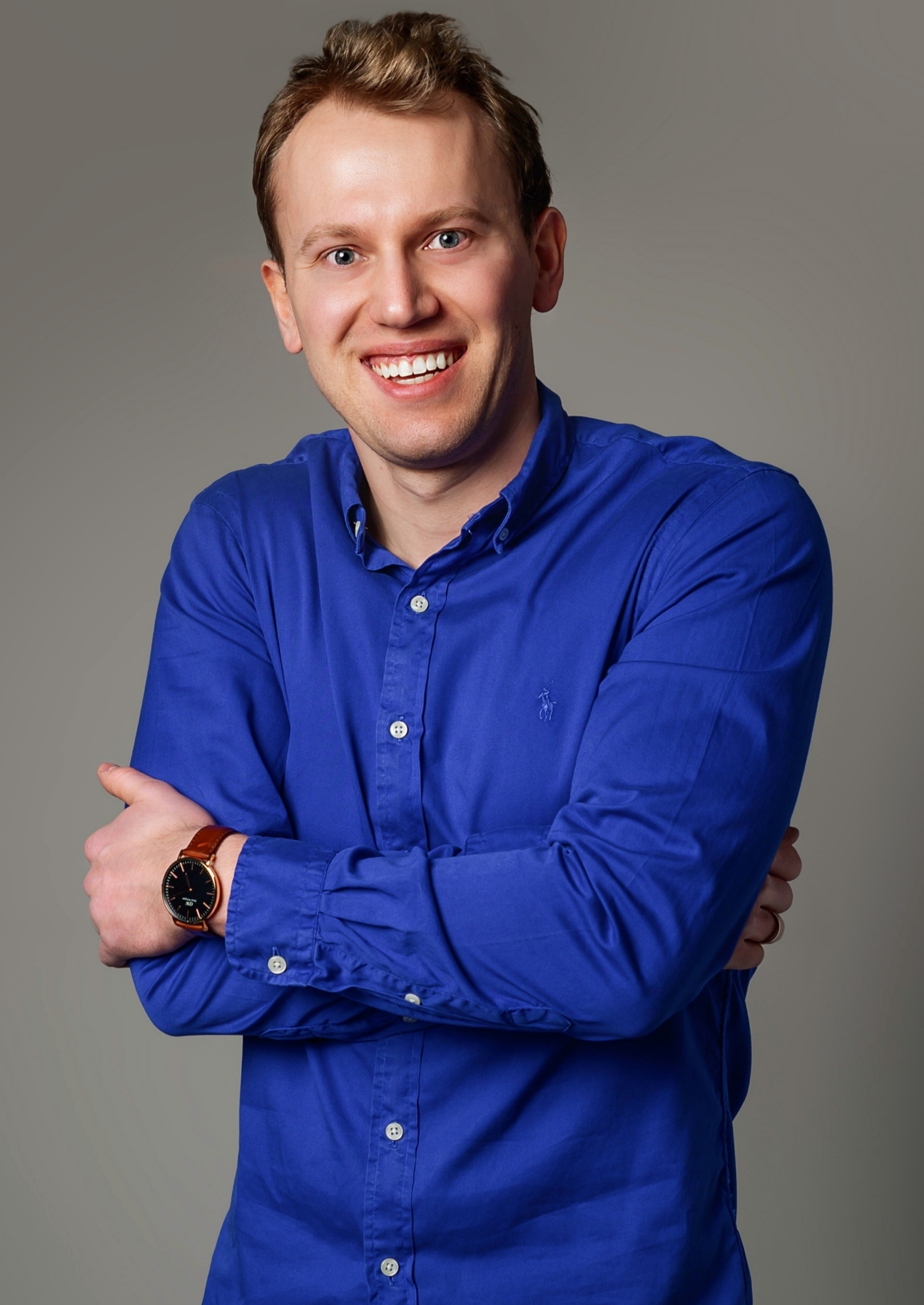}}]{Artur Laskowski} is a specialist with 9 years of commercial experience in developing and implementing systems serving millions of users. Lecturer at Lodz University of Technology and an IT trainer, sharing knowledge and practical experience.
\end{IEEEbiography}

\vfill

\end{document}